\RequirePackage{lineno}
\documentclass[aps,prd,twocolumn,superscriptaddress,showpacs,amsmath,amssymb]{revtex4}
\usepackage{comment}
\usepackage{multirow}
\usepackage{dcolumn}% Align table columns on decimal point
\usepackage{bm}% bold math
\usepackage{setspace}
\usepackage{graphicx}% Include figure files
\includecomment{printrobustnotes}
\includecomment{printallnotes}
\usepackage{xspace}

\usepackage{dcolumn}% Align table columns on decimal point
\usepackage{bm}% bold math
\newenvironment{cfigure1c}[1][tbp]{\begin{figure*}[#1]\centering}{\end{figure*}}

\newcommand{\note}[1]{\xspace}
\newcommand{\robustnote}[1]{\xspace}
\begin{printrobustnotes}
\renewcommand{\robustnote}[1]{\textbf{\textit{#1}}\xspace}
\end{printrobustnotes}
\begin{printallnotes}
\renewcommand{\note}[1]{\textbf{\textit{#1}}\xspace}
\renewcommand{\robustnote}[1]{\textbf{\textit{#1}}\xspace}
\end{printallnotes}
\newcommand{\ppbar}  {\ensuremath{p\bar{p}}\xspace}
\newcommand{\ttbar}  {\ensuremath{t\bar{t}}\xspace}

\newcommand{\Ht}     {\ensuremath{H_{T}}\xspace}
\newcommand{\et}     {\ensuremath{E_{T}}\xspace}
\newcommand{\etl}     {\ensuremath{E_{T}^{\mathrm{lepton}}}\xspace}
\newcommand{\etj}     {\ensuremath{E_{T}^{\mathrm{jet}}}\xspace}

\newcommand{\pt}     {\ensuremath{p_{T}}\xspace}

\newcommand{\met}    {\ensuremath{\text{\raisebox{.3ex}{$\not$}}\et}\xspace}

\newcommand{\dmt}{\ensuremath{\Delta\mathrm{M}_{\text{top}}}\xspace}
\newcommand{\dmtr}{\ensuremath{\Delta m_{\text{reco}}}\xspace}
\newcommand{\dmttr}{\ensuremath{\Delta m_{\text{reco}}^{(2)}}\xspace}

\newcommand{\chisq}  {\ensuremath{\chi^{2}}\xspace}

\newcommand{\mtop}    {\ensuremath{\mathrm{M}_{\text{top}}}\xspace}
\newcommand{\genunit}[2]{\ensuremath{#1~\mathrm{#2}}\xspace}

\newcommand{\gev}[1]    {\genunit{#1}{GeV}}
\newcommand{\pb}[1]     {\genunit{#1}{pb}}

\newcommand{\gevc}[1]   {\ensuremath{#1~\mathrm{GeV}/c}}
\newcommand{\gevcc}[1]  {\ensuremath{#1~\mathrm{GeV}/c^{2}}}

\newcommand{\invfb}[1]  {\ensuremath{#1~\mathrm{fb}^{-1}}}

\newcommand{\gevccnoarg}{\ensuremath{\mathrm{GeV}/c^{2}}\xspace}

\newcommand{\measErr}[2]{\ensuremath{#1 \pm #2}\xspace}

\newcommand{\measStatSyst}[3]{\ensuremath{#1 \pm #2~(\textrm{stat}) \pm #3~(\textrm{syst})}\xspace}

\begin{document}
%\setpagewiselinenumbers
%\modulolinenumbers[1]
%\linenumbers

\title{Measurement of the mass difference between $t$ and $\bar{t}$ quarks}

\affiliation{Institute of Physics, Academia Sinica, Taipei, Taiwan 11529, Republic of China} 
\affiliation{Argonne National Laboratory, Argonne, Illinois 60439, USA} 
\affiliation{University of Athens, 157 71 Athens, Greece} 
\affiliation{Institut de Fisica d'Altes Energies, ICREA, Universitat Autonoma de Barcelona, E-08193, Bellaterra (Barcelona), Spain} 
\affiliation{Baylor University, Waco, Texas 76798, USA} 
\affiliation{Istituto Nazionale di Fisica Nucleare Bologna, $^z$University of Bologna, I-40127 Bologna, Italy} 
\affiliation{University of California, Davis, Davis, California 95616, USA} 
\affiliation{University of California, Los Angeles, Los Angeles, California 90024, USA} 
\affiliation{Instituto de Fisica de Cantabria, CSIC-University of Cantabria, 39005 Santander, Spain} 
\affiliation{Carnegie Mellon University, Pittsburgh, Pennsylvania 15213, USA} 
\affiliation{Enrico Fermi Institute, University of Chicago, Chicago, Illinois 60637, USA}
\affiliation{Comenius University, 842 48 Bratislava, Slovakia; Institute of Experimental Physics, 040 01 Kosice, Slovakia} 
\affiliation{Joint Institute for Nuclear Research, RU-141980 Dubna, Russia} 
\affiliation{Duke University, Durham, North Carolina 27708, USA} 
\affiliation{Fermi National Accelerator Laboratory, Batavia, Illinois 60510, USA} 
\affiliation{University of Florida, Gainesville, Florida 32611, USA} 
\affiliation{Laboratori Nazionali di Frascati, Istituto Nazionale di Fisica Nucleare, I-00044 Frascati, Italy} 
\affiliation{University of Geneva, CH-1211 Geneva 4, Switzerland} 
\affiliation{Glasgow University, Glasgow G12 8QQ, United Kingdom} 
\affiliation{Harvard University, Cambridge, Massachusetts 02138, USA} 
\affiliation{Division of High Energy Physics, Department of Physics, University of Helsinki and Helsinki Institute of Physics, FIN-00014, Helsinki, Finland} 
\affiliation{University of Illinois, Urbana, Illinois 61801, USA} 
\affiliation{The Johns Hopkins University, Baltimore, Maryland 21218, USA} 
\affiliation{Institut f\"{u}r Experimentelle Kernphysik, Karlsruhe Institute of Technology, D-76131 Karlsruhe, Germany} 
\affiliation{Center for High Energy Physics: Kyungpook National University, Daegu 702-701, Korea; Seoul National University, Seoul 151-742, Korea; Sungkyunkwan University, Suwon 440-746, Korea; Korea Institute of Science and Technology Information, Daejeon 305-806, Korea; Chonnam National University, Gwangju 500-757, Korea; Chonbuk National University, Jeonju 561-756, Korea} 
\affiliation{Ernest Orlando Lawrence Berkeley National Laboratory, Berkeley, California 94720, USA} 
\affiliation{University of Liverpool, Liverpool L69 7ZE, United Kingdom} 
\affiliation{University College London, London WC1E 6BT, United Kingdom} 
\affiliation{Centro de Investigaciones Energeticas Medioambientales y Tecnologicas, E-28040 Madrid, Spain} 
\affiliation{Massachusetts Institute of Technology, Cambridge, Massachusetts 02139, USA} 
\affiliation{Institute of Particle Physics: McGill University, Montr\'{e}al, Qu\'{e}bec, Canada H3A~2T8; Simon Fraser University, Burnaby, British Columbia, Canada V5A~1S6; University of Toronto, Toronto, Ontario, Canada M5S~1A7; and TRIUMF, Vancouver, British Columbia, Canada V6T~2A3} 
\affiliation{University of Michigan, Ann Arbor, Michigan 48109, USA} 
\affiliation{Michigan State University, East Lansing, Michigan 48824, USA}
\affiliation{Institution for Theoretical and Experimental Physics, ITEP, Moscow 117259, Russia}
\affiliation{University of New Mexico, Albuquerque, New Mexico 87131, USA} 
\affiliation{Northwestern University, Evanston, Illinois 60208, USA} 
\affiliation{The Ohio State University, Columbus, Ohio 43210, USA} 
\affiliation{Okayama University, Okayama 700-8530, Japan} 
\affiliation{Osaka City University, Osaka 588, Japan} 
\affiliation{University of Oxford, Oxford OX1 3RH, United Kingdom} 
\affiliation{Istituto Nazionale di Fisica Nucleare, Sezione di Padova-Trento, $^{aa}$University of Padova, I-35131 Padova, Italy} 
\affiliation{LPNHE, Universite Pierre et Marie Curie/IN2P3-CNRS, UMR7585, Paris, F-75252 France} 
\affiliation{University of Pennsylvania, Philadelphia, Pennsylvania 19104, USA}
\affiliation{Istituto Nazionale di Fisica Nucleare Pisa, $^{bb}$University of Pisa, $^{cc}$University of Siena and $^{dd}$Scuola Normale Superiore, I-56127 Pisa, Italy} 
\affiliation{University of Pittsburgh, Pittsburgh, Pennsylvania 15260, USA} 
\affiliation{Purdue University, West Lafayette, Indiana 47907, USA} 
\affiliation{University of Rochester, Rochester, New York 14627, USA} 
\affiliation{The Rockefeller University, New York, New York 10065, USA} 
\affiliation{Istituto Nazionale di Fisica Nucleare, Sezione di Roma 1, $^{ee}$Sapienza Universit\`{a} di Roma, I-00185 Roma, Italy} 

\affiliation{Rutgers University, Piscataway, New Jersey 08855, USA} 
\affiliation{Texas A\&M University, College Station, Texas 77843, USA} 
\affiliation{Istituto Nazionale di Fisica Nucleare Trieste/Udine, I-34100 Trieste, $^{ff}$University of Trieste/Udine, I-33100 Udine, Italy} 
\affiliation{University of Tsukuba, Tsukuba, Ibaraki 305, Japan} 
\affiliation{Tufts University, Medford, Massachusetts 02155, USA} 
\affiliation{University of Virginia, Charlottesville, VA  22906, USA}
\affiliation{Waseda University, Tokyo 169, Japan} 
\affiliation{Wayne State University, Detroit, Michigan 48201, USA} 
\affiliation{University of Wisconsin, Madison, Wisconsin 53706, USA} 
\affiliation{Yale University, New Haven, Connecticut 06520, USA} 
\author{T.~Aaltonen}
\affiliation{Division of High Energy Physics, Department of Physics, University of Helsinki and Helsinki Institute of Physics, FIN-00014, Helsinki, Finland}
\author{B.~\'{A}lvarez~Gonz\'{a}lez$^v$}
\affiliation{Instituto de Fisica de Cantabria, CSIC-University of Cantabria, 39005 Santander, Spain}
\author{S.~Amerio}
\affiliation{Istituto Nazionale di Fisica Nucleare, Sezione di Padova-Trento, $^{aa}$University of Padova, I-35131 Padova, Italy} 

\author{D.~Amidei}
\affiliation{University of Michigan, Ann Arbor, Michigan 48109, USA}
\author{A.~Anastassov}
\affiliation{Northwestern University, Evanston, Illinois 60208, USA}
\author{A.~Annovi}
\affiliation{Laboratori Nazionali di Frascati, Istituto Nazionale di Fisica Nucleare, I-00044 Frascati, Italy}
\author{J.~Antos}
\affiliation{Comenius University, 842 48 Bratislava, Slovakia; Institute of Experimental Physics, 040 01 Kosice, Slovakia}
\author{G.~Apollinari}
\affiliation{Fermi National Accelerator Laboratory, Batavia, Illinois 60510, USA}
\author{J.A.~Appel}
\affiliation{Fermi National Accelerator Laboratory, Batavia, Illinois 60510, USA}
\author{A.~Apresyan}
\affiliation{Purdue University, West Lafayette, Indiana 47907, USA}
% ...
\collaboration{CDF Collaboration\footnote{With visitors from $^a$University of Massachusetts Amherst, Amherst, Massachusetts 01003,
$^b$Istituto Nazionale di Fisica Nucleare, Sezione di Cagliari, 09042 Monserrato (Cagliari), Italy,
$^c$University of California Irvine, Irvine, CA  92697, 
$^d$University of California Santa Barbara, Santa Barbara, CA 93106
$^e$University of California Santa Cruz, Santa Cruz, CA  95064,
$^f$CERN,CH-1211 Geneva, Switzerland,
$^g$Cornell University, Ithaca, NY  14853, 
$^h$University of Cyprus, Nicosia CY-1678, Cyprus, 
$^i$University College Dublin, Dublin 4, Ireland,
$^j$University of Fukui, Fukui City, Fukui Prefecture, Japan 910-0017,
$^k$Universidad Iberoamericana, Mexico D.F., Mexico,
$^l$Iowa State University, Ames, IA  50011,
$^m$University of Iowa, Iowa City, IA  52242,
$^n$Kinki University, Higashi-Osaka City, Japan 577-8502,
$^o$Kansas State University, Manhattan, KS 66506,
$^p$University of Manchester, Manchester M13 9PL, England,
$^q$Queen Mary, University of London, London, E1 4NS, England,
$^r$Muons, Inc., Batavia, IL 60510,
$^s$Nagasaki Institute of Applied Science, Nagasaki, Japan, 
$^t$National Research Nuclear University, Moscow, Russia,
$^u$University of Notre Dame, Notre Dame, IN 46556,
$^v$Universidad de Oviedo, E-33007 Oviedo, Spain, 
$^w$Texas Tech University, Lubbock, TX  79609, 
$^x$Universidad Tecnica Federico Santa Maria, 110v Valparaiso, Chile,
$^y$Yarmouk University, Irbid 211-63, Jordan,
$^{gg}$On leave from J.~Stefan Institute, Ljubljana, Slovenia, 
}}
\noaffiliation

\date{\today}% It is always \today, today,
\begin{abstract}
We present a direct measurement of the mass difference between $t$ and $\bar{t}$ quarks using $\ttbar$
candidate events in the lepton+jets channel, collected with the CDF II detector at Fermilab's 1.96~TeV Tevatron 
\ppbar Collider. We make an event by event estimate of the mass difference to construct templates for top quark pair signal events and 
background events. The resulting mass difference distribution of data is compared to templates of signals and background using a maximum likelihood fit.
From a sample corresponding to an integrated luminosity of \invfb{5.6}, we measure a mass difference, 
$\dmt = \mathrm{M}_{t} - \mathrm{M}_{\bar{t}} = \gevcc{\measStatSyst{-3.3}{1.4}{1.0}}$, 
approximately two standard deviations away from the CPT hypothesis of zero mass difference. 
%This is the most precise measurement of a mass difference between a quark and its anti-quark partner to date. 

\end{abstract}

% PACS, high energy physics
\maketitle

Discrete symmetries reflecting the invariance under discrete transformations, such as charge conjugation~(C), 
space reflection or parity~(P), and time reversal~(T), are not always exact. 
Examples include the C and P symmetries and their CP combination, which are violated by the weak interactions~\cite{pdg}. 
CPT symmetry, which reflects the invariance under the combined operation 
of C, P, and T transformations, has not been found to be violated in any experiment so far~\cite{kaoncpt, nucpt}. However,
 it is important to examine the possibility of CPT violation in all sectors of the standard model (SM), as there are 
well-motivated extensions of the SM allowing for  CPT symmetry breaking~\cite{cptv}.  
In the CPT theorem, particle and anti-particle masses must be identical; thus,
a mass difference between particle and its anti-particle would indicate a violation of CPT. 
The mass equality has been verified to high precision for leptons and hadrons, but not for quarks. 
With the exception of the top quark, it is impossible to measure quark masses directly, because a newly created quark 
dresses itself with other quarks and gluons to form a hadron, and 
hadron masses yield, at best, only rough estimates of the quark mass. 
The top quark is by far the most massive quark and, with lifetime of 
the order of $10^{-24}$ seconds, decays before it can hadronize. This allows a precise measurement of the mass difference
between $t$ and $\bar{t}$ quarks and provides a probe of CPT violation in the quark sector~\cite{cpttop}.

This letter reports a measurement of the mass difference~($\dmt = \mathrm{M}_{t} -\mathrm{M}_{\bar{t}}$) 
between $t$ and $\bar{t}$ quarks using a sample of \ttbar candidates in the lepton+jets 
final state. 
The data correspond to an integrated luminosity of \invfb{5.6} in proton-antiproton collisions at the Tevatron with $\sqrt{s}=1.96$~TeV,
collected with the CDF II detector~\cite{r_cdfDetector}. 
Assuming unitarity of the three-generation CKM matrix, $t$ and $\bar{t}$ quarks decay almost exclusively into a $W$ boson and a bottom quark ($t\rightarrow bW^{+}$ and $\bar{t}\rightarrow \bar{b}W^{-}$)~\cite{pdg}. 
The case where one $W$ decays into a charged lepton and a neutrino ($W^{+}\rightarrow \bar{\ell}\nu$ or $W^{-}\rightarrow \ell\bar{\nu}$) and 
the other into a pair of jets defines the lepton+jets decay channel. The electric charge of the lepton~(-1 for $\ell$ and +1 for $\bar{\ell}$) determines the flavor of top quarks with event reconstruction. 
To select $\ttbar$ candidate events in this channel, we require one electron (muon) with 
$\et > \gev{20}$ ($\pt > \gevc{20}$) and pseudorapidity $|\eta|<1.1$~\cite{cdfco}.  
We also require high missing transverse energy~\cite{met}, $\met > \gev{20}$, and at least four jets. 
Jets are reconstructed with a cone algorithm~\cite{jetclu} with radius 
$R = \sqrt{(\Delta \eta)^2+(\Delta \phi)^2} = 0.4$.
Jets originating from 
$b$ quarks are identified using a secondary vertex tagging algorithm~\cite{secvtx}. 
In order to optimize the background reduction process and improve the statistical power of the events, we divide the sample of \ttbar candidate events into sub-samples with zero, one, and two or more $b$-tagged jets. 

When an event has zero or one $b$-tagged jet, we require exactly four jets 
with transverse energy $\et > \gev{20}$ and $|\eta|<2.0$. If an event has two or more $b$-jets, three jets are 
required to have $\et > \gev{20}$ and $|\eta| <2.0$, and a fourth jet is required to have $\et > \gev{12}$ 
and $|\eta|<2.4$, with no restriction on the total number of jets. 
To reject backgrounds, we require the scalar sum of transverse energies in the event, 
$\Ht = \etl + \met + \sum_{\text{four jets}}\etj$, to be greater than 250~GeV.

\begin{table}
\begin{center}
\caption[Expected number of background and signal events]{
Expected and observed numbers of signal and background events 
assuming \ttbar production cross-section $\sigma_{\ttbar}$ = \pb{7.4} and \mtop = \gevcc{172.5}. 
}
\begin{ruledtabular}
\label{table_background}
\begin{tabular}{lccc}
             &0 $b$-tag        &1 $b$-tag       &$\geq$ 2 $b$-tag\\
\hline
W+jets       &596 $\pm$ 98 &88.3 $\pm$ 23.0 &11.1 $\pm$ 3.6\\
QCD multijet & 95.8 $\pm$ 74.4 &14.7 $\pm$ 12.1 & 2.4 $\pm$ 3.2\\
Z+jets       & 48.8 $\pm$ 9.4  & 5.7 $\pm$ 1.3  & 0.8 $\pm$ 0.2\\
Diboson      & 50.1 $\pm$ 4.7  & 6.6 $\pm$ 0.8  & 1.0 $\pm$ 0.2\\
Single top   &  4.0 $\pm$ 0.4  & 5.5 $\pm$ 0.5  & 2.2 $\pm$ 0.2\\
\hline
Background   & 795$\pm$ 124 &121 $\pm$ 24 & 17.3 $\pm$ 4.8 \\
\ttbar signal&426 $\pm$ 57 &578 $\pm$ 72 & 282 $\pm$ 44\\
\hline
Expected& 1220 $\pm$ 137 & 699 $\pm$ 76  & 299 $\pm$ 44 \\
Observed     &1278 & 720 & 296 \\
\end{tabular}
\end{ruledtabular}
\end{center}
\end{table}

The primary sources of background events are $W$+jets and QCD multijet production. Contributions
from $Z$+jets, diboson, and single top production are expected to be small.
To estimate the contribution of each process, we use a combination of data and Monte Carlo~(MC) 
based techniques described in Ref.~\cite{xsection}. For the $Z$+jets, diboson, and single top quark events, 
we normalized MC simulation events using their respective theoretical cross sections. The QCD multijet background is estimated with a 
data-driven approach. We model $W$+jets background events using MC simulation but the overall rate is determined using data after subtracting the rate of all the other backgrounds and \ttbar. 
Table~\ref{table_background} shows the expected background composition and the expected number of \ttbar events. 

We assume selected events to be $\ttbar$ events in the lepton+jets channel and reconstruct them 
to form estimators of \dmt, using a special purpose kinematic fitter, in which we modify the standard fitter~\cite{fitter} to allow a mass difference between $t$ and $\bar{t}$. 
Measured four-vectors of jets and lepton are corrected for known effects~\cite{jes}, 
and resolutions are assigned. The unclustered transverse energy~($U_{T}$), which is the sum of all transverse energy in the calorimeter that is not associated with the primary lepton or one of the leading four jets, is used to calculate the neutrino transverse momentum. The longitudinal momentum of the neutrino is a free~(unconstrained) parameter which is effectively determined by the constraint on the invariant mass of the leptonic $W$. 
We then define a kinematic fit $\chi^{2}$ having a free parameter $dm_{reco}$, 
\begin{linenomath*}
\begin{eqnarray}
\label{eq_chi2}
\chisq & = &
\Sigma_{i = \ell, 4 jets} {(p_T^{i,fit} - p_T^{i,meas})^2 / \sigma_i^2} \nonumber \\
& + & \Sigma_{k = x,y} {(U_{T_k}^{fit} - U_{T_k}^{meas})^2 / \sigma_k^2} \nonumber \\
& + & {(\mathrm{M}_{jj} - \mathrm{M}_W)^2 / \Gamma_W^2} 
 + {(\mathrm{M}_{\ell\nu} - \mathrm{M}_W)^2 / \Gamma_W^2} \nonumber \\
& + &  {\{\mathrm{M}_{bjj} - ({\overline{\mathrm M}}_{\text{top}}+{dm}_{reco}/2)\}^2 / \Gamma_t^2} \nonumber \\
& + & {\{\mathrm{M}_{b\ell\nu} - ({\overline{\mathrm M}}_{\text{top}}-{dm}_{reco}/2)\}^2 / \Gamma_t^2},
%& + &  {(\mathrm{M}_{bjj} - m_{reco})^2 / \Gamma_t^2} \nonumber \\
%& + & {(\mathrm{M}_{b\ell\nu} - m_{reco})^2 / \Gamma_t^2},
\end{eqnarray}
\end{linenomath*}
where $dm_{reco}^{min}$, the $dm_{reco}$ value at the lowest $\chi^2$, represents 
the reconstructed mass difference between the hadronic and leptonic top 
decay~($\mathrm{M}_{bjj}-\mathrm{M}_{b\ell\nu}$). 
In this $\chi^2$ formulation, the first term constrains the \pt of the lepton and 
four leading jets to their measured 
values within their uncertainties~($\sigma_i$); the second term does the same for both transverse 
components $x$ and $y$ of the unclustered transverse energy. In the remaining four terms, the quantities 
$\mathrm{M}_{jj}, \mathrm{M}_{\ell\nu}, \mathrm{M}_{bjj}$, and $\mathrm{M}_{b\ell\nu}$ 
refer to the invariant masses of the four vector sum of the particles denoted in the subscripts. 
$\mathrm{M}_W$ and ${\overline{\mathrm M}}_{\text{top}}$ are the masses of the $W$ boson~(\gevcc{80.4})~\cite{pdg} and 
the average of $t$ and $\bar t$ quark masses~(\gevcc{172.5}), close to the current best experimental determination~\cite{mtop},
respectively. 
$\Gamma_W$ (\gevcc{2.1}) and $\Gamma_t$ (\gevcc{1.5}) are the total widths of the $W$ boson 
and the $t$ quark~\cite{pdg}. We assume that the total widths of the $t$ and $\bar t$ quarks are equal.  
Determining the reconstructed mass difference of $t$ and $\bar{t}$, $\dmtr$, requires the
identification of the flavor ($t$ versus $\bar t$), and this is done using the electric charge
of the lepton~($Q_{lepton}$), defining 
$\dmtr = -Q_{lepton} \times {dm}_{reco}^{min}$.

The use of different detector components and the different resolutions of the measured values for jet, 
lepton, and unclustered energy, make the reconstructed mass distribution of hadronic 
top quarks differ from that of leptonic top quarks. Because the sign of $\dmtr$  
depends on the lepton charge, $\dmtr$ distributions for the positive and negative lepton events
are different. We divide the sample into six sub-samples, two samples with positively and negatively 
charged leptons for each of 0 $b$-tag, 1 $b$-tag, and 2 $b$-tag samples.

With the assumption that the leading four jets in the event come from the four final 
quarks at the hard scattering level, there are 12, 6, and 2 possible assignments of jets to quarks for 0 $b$-tag, 1 $b$-tag, and 2 $b$-tag respectively. 
The minimization of $\chi^2$ is performed for each jet-to-parton assignment, and \dmtr is taken from the assignment 
that yields the lowest $\chi^2$ ($\chi^2_{min}$). Events with 
$\chi^2_{min} > 9.0$~($\chi^2_{min} > 3.0$) 
are removed from the sample to reject poorly reconstructed events for $b$-tagged~(zero 
$b$-tagged) events. To increase the statistical power of the measurement, we employ an 
additional observable \dmttr from the assignment that yields the 2$^{nd}$ lowest $\chi^2$. Although it has a poorer sensitivity, \dmttr 
provides additional information on \dmt and improves the statistical 
uncertainty by approximately 10~\%.

Using {\sc madgraph}~\cite{madgraph}, we generate \ttbar signal samples
with \dmt between \gevcc{-20} and \gevcc{20} using almost \gevcc{2} step size, where we take the average mass value of $t$ and $\bar t$ to 
be ${\overline{\mathrm M}}_{\text{top}} = \gevcc{172.5}$. Parton showering of the signal events is simulated with 
{\sc pythia}~\cite{pythia}, and the CDF detector is simulated using a {\sc geant}-based 
software package~\cite{cdfsim}. 

We estimate the probability density functions~(p.d.f.s) of signal and
background templates using the kernel density estimation~(KDE)~\cite{KDEHEP,tmt19}.
For the \dmt measurement with two observables~(\dmtr and \dmttr), we use the two 
dimensional KDE that accounts for the correlation between them.
First, at discrete values of \dmt from \gevcc{-20} to \gevcc{20}, we estimate 
the p.d.f.s for the observables from above-mentioned $\ttbar$ MC samples. 
We interpolate the MC distributions 
to find p.d.f.s for arbitrary values of \dmt using the local polynomial smoothing
method~\cite{lps}.  We fit the signal and background p.d.f.s to the measured distributions 
of the observables in the data using an unbinned maximum likelihood fit~\cite{eml}, where
we minimize the negative logarithm of the likelihood with {\sc minuit}~\cite{minuit}. 
Likelihoods are built for each of six sub-samples separately, and an overall likelihood is then obtained by multiplying them together. 
We evaluate the statistical uncertainty on \dmt by searching for the points where the
negative logarithm of the likelihood 
exceeds the minimum by 0.5.
Refs.~\cite{tmt19,tmtmt2} provide detailed information about this
technique.

We test the fitting procedure using 3000 MC pseudo experiments~(PEs) 
for each of 11 equally spaced $\dmt$ values ranging from \gevcc{-10} to \gevcc{10}. 
The distributions of the average residual of measured \dmt~(deviation from the input \dmt) 
for simulated experiments is consistent with zero.
However the width of the pull~(the ratio of the residual to the uncertainty reported 
by {\sc minuit}) is 4~\% greater than unity. We therefore increase the measured uncertainty by 4~\%. 

\begin{table}
\begin{center}
\caption[Summary of systematics]{Summary of systematic uncertainties on \dmt.}
\begin{ruledtabular}
\label{systtablesummary}
\begin{tabular}{lc}
Source              & Uncertainty~(\gevccnoarg) \\
\hline
Signal modeling                    & 0.7 \\
$b$ and $\bar{b}$ jets asymmetry        & 0.4 \\
Jet energy scale                   & 0.2 \\
Parton distribution functions      & 0.1 \\
$b$-jet energy scale               & 0.1 \\
Background shape                   & 0.2 \\
Gluon fusion fraction              & 0.1 \\
Initial and final state radiation  & 0.1 \\
Monte Carlo statistics             & 0.1 \\
Lepton energy scale                & 0.1 \\
Multiple hadron interaction        & 0.4 \\
Color reconnection                 & 0.2 \\
\hline
Total systematic uncertainty       & 1.0 \\
\end{tabular}
\end{ruledtabular}
\end{center}
\end{table}

We examine a variety of systematic effects that could change the measurement
by comparing results from PEs in which we vary relevant systematic
parameters within their uncertainties. All systematic uncertainties are summarized 
in Table~\ref{systtablesummary}. The dominant source of systematic uncertainty 
is the signal modeling, which we estimate using PEs with events generated with {\sc madgraph} 
and {\sc pythia}. We also estimate a parton showering uncertainty by applying different showering models 
({\sc pythia} and {\sc herwig}~\cite{herwig}) to a sample generated with {\sc alpgen}~\cite{alpgen}. 
We address a possible difference in the detector response 
between $b$ and $\bar{b}$ jets by comparing data and MC simulation events~\cite{bbasym}. 
We add a systematic uncertainty due to multiple hadron interactions to account 
for the fact that the average number of interactions in our MC samples is not exactly equal to the 
number observed in the data. 
The jet energy scale~(JES), the dominant uncertainty in most of the top quark mass measurements, is partially canceled in the measurement of the mass difference. Therefore JES contributes only a small uncertainty to this measurement. 
Other sources of systematic effects, 
including uncertainties in parton distribution functions, gluon radiation, background shape and 
normalization, lepton energy scale, and color reconnection~\cite{CR,tmtmt2}, give small contributions.
The total systematic uncertainty of \gevcc{1.0} is derived from a quadrature sum of the listed uncertainties. 
 
\begin{cfigure1c}
\begin{tabular}{cc}
\includegraphics[width=0.43\textwidth]{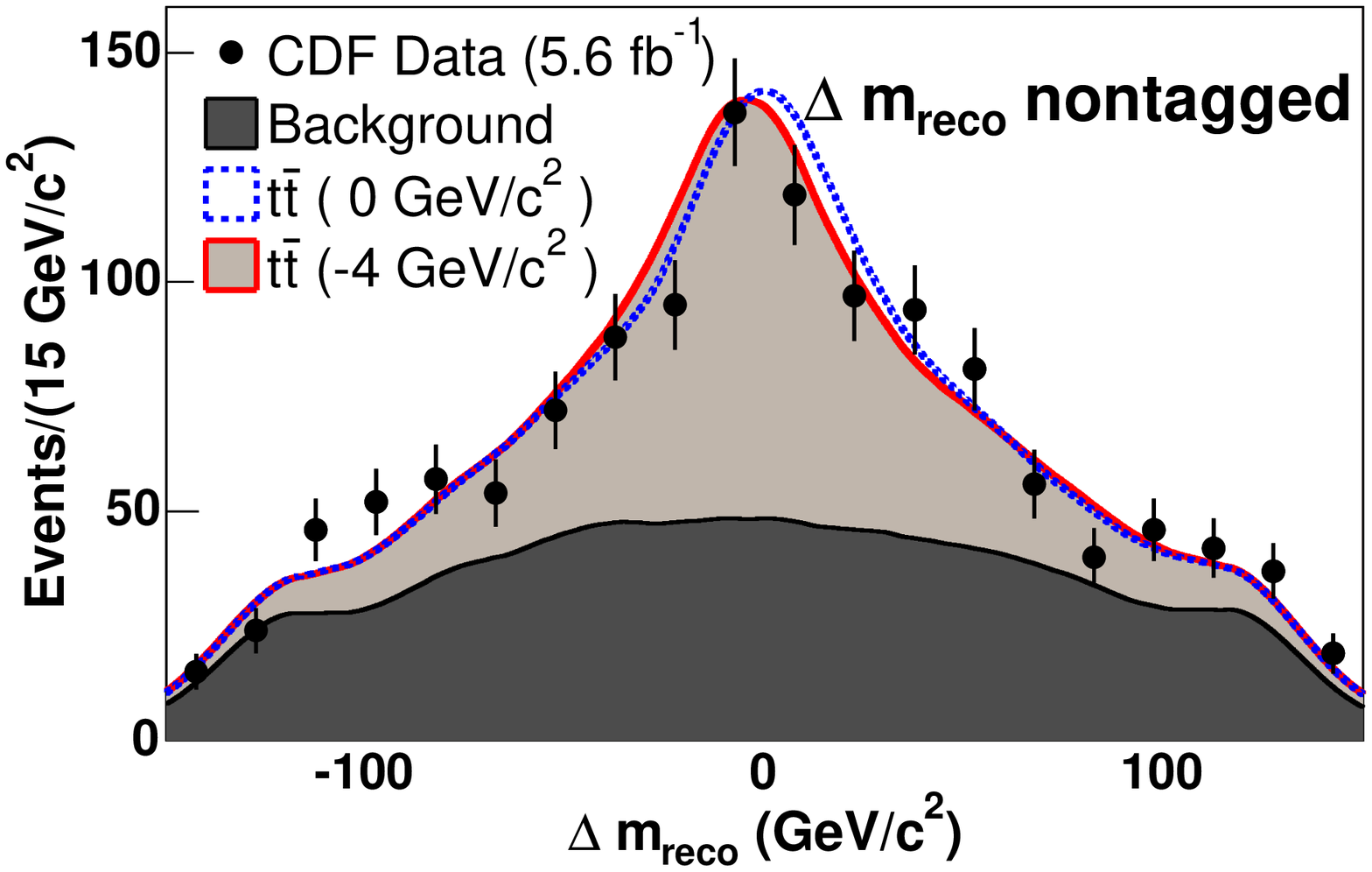} &
\includegraphics[width=0.43\textwidth]{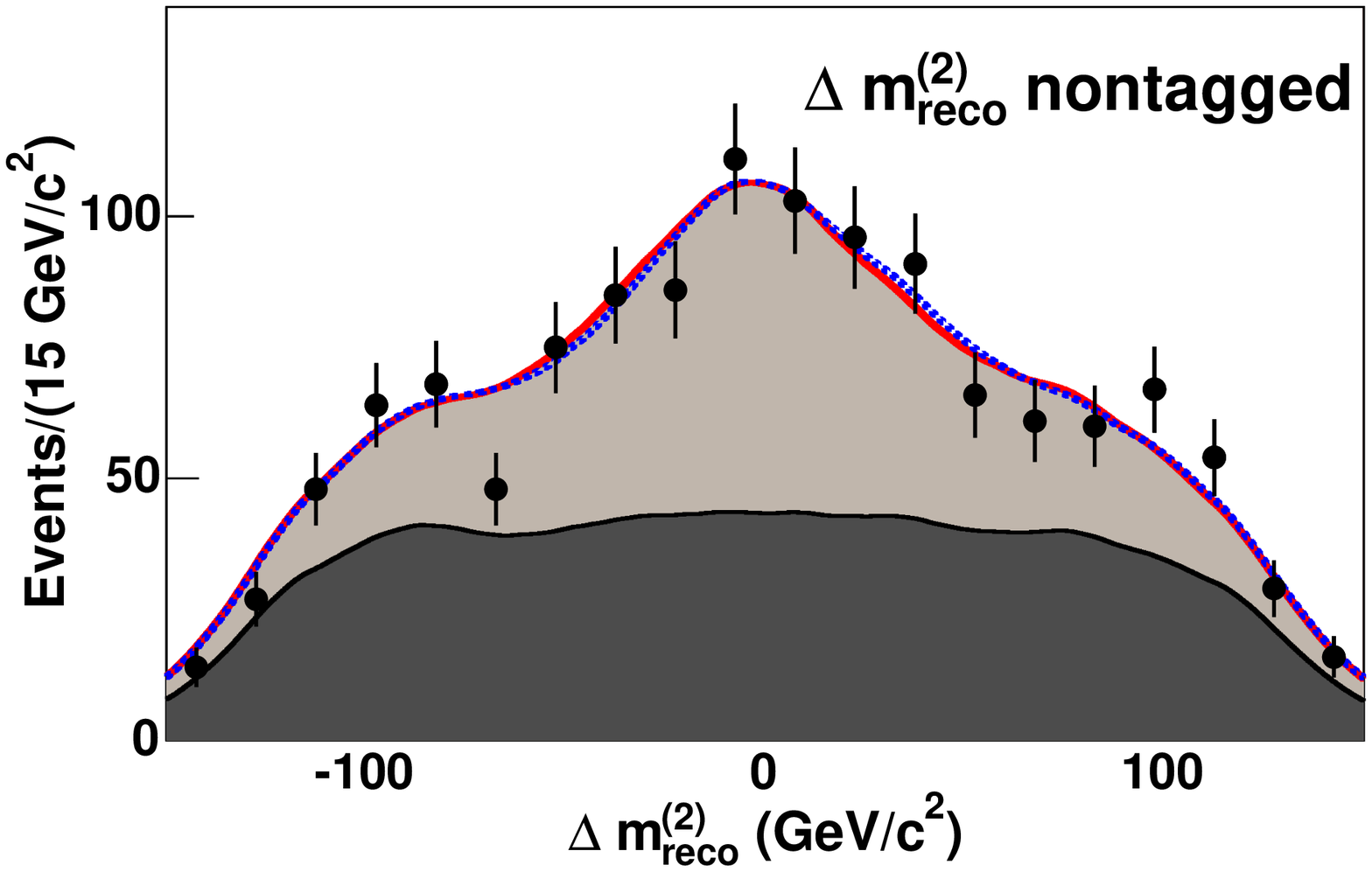} \\
\includegraphics[width=0.43\textwidth]{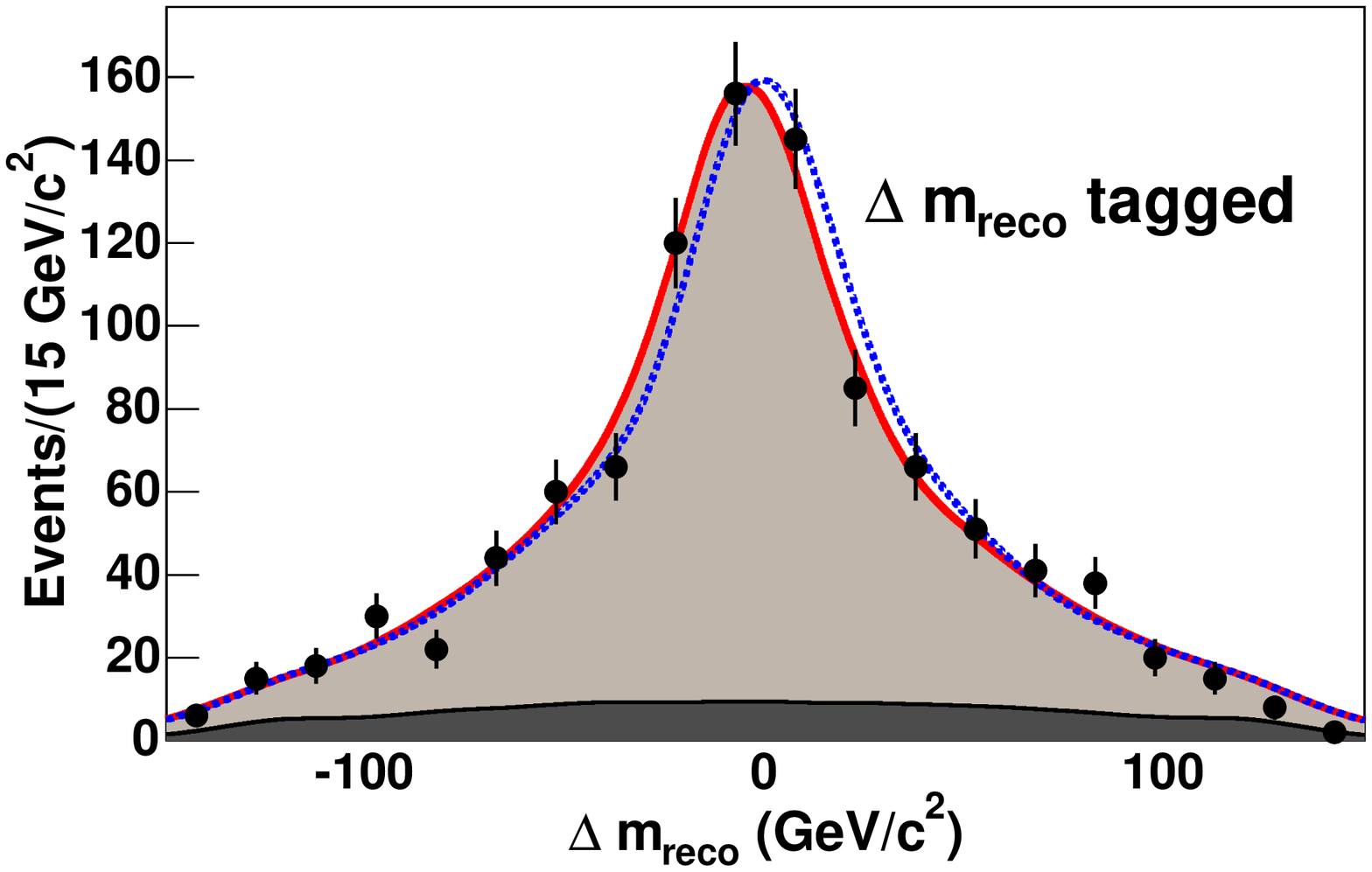} &
\includegraphics[width=0.43\textwidth]{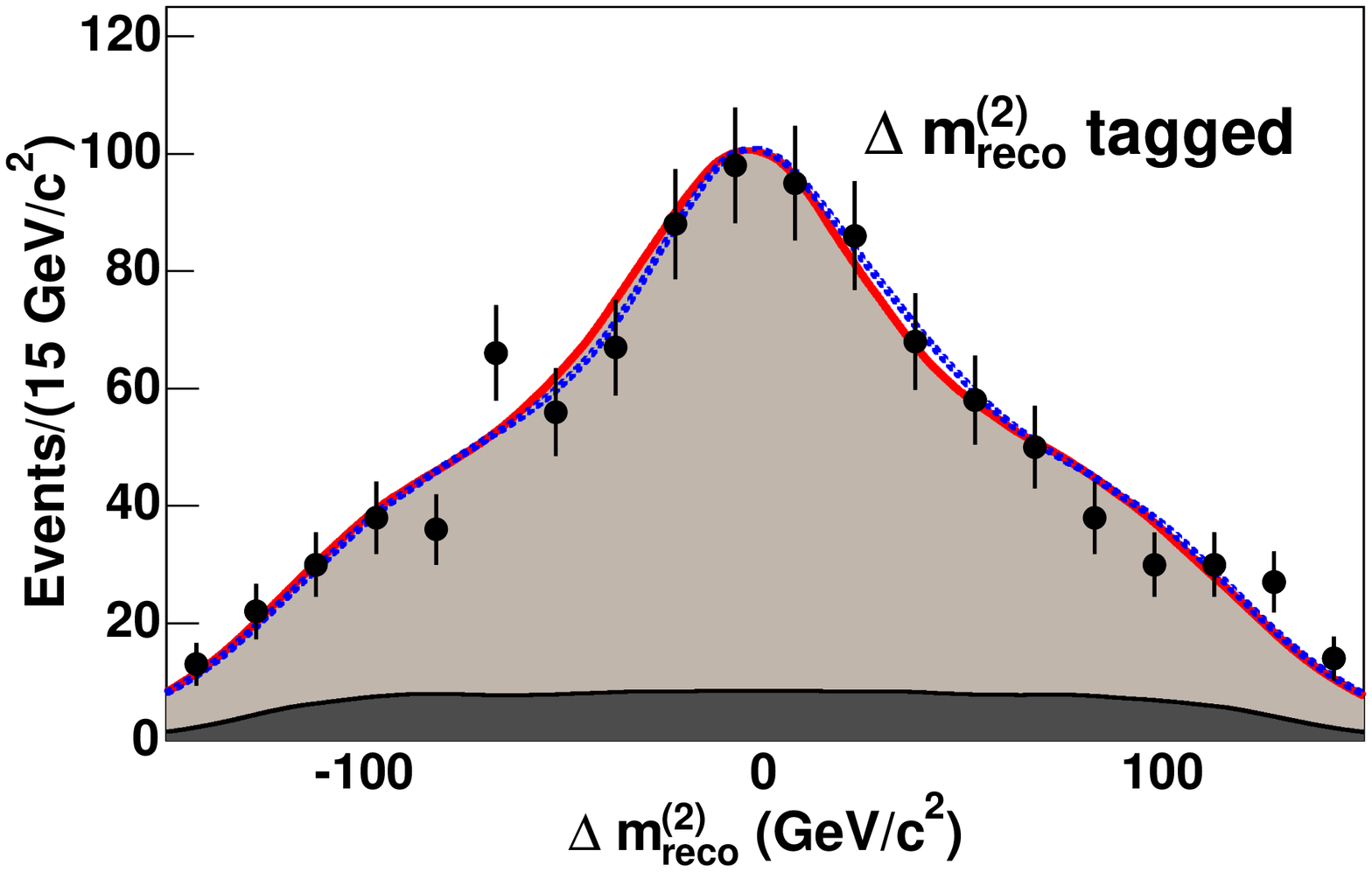} \\
\end{tabular}
\caption[Data fit]{(Color online) Distributions of $\dmtr$ and $\dmttr$ used to extract \dmt for zero 
$b$-tagged~(nontagged) events and one or more $b$-tagged~(tagged)  events. 
The data is overlaid with the predictions from the KDE probability 
distributions assuming \dmt~=~\gevcc{-4}~(solid red line) and \dmt~=~\gevcc{0}~(dashed blue line).
}
\label{ref:fit}
\end{cfigure1c}

The likelihood fit to the data returns a mass difference
\begin{linenomath*}
%\begin{align*}
\begin{eqnarray*}
\dmt & = \gevcc{\measStatSyst{-3.3}{1.4}{1.0}}\\ 
     & = \gevcc{\measErr{-3.3}{1.7}}.
\end{eqnarray*}
%\end{align*}
\end{linenomath*} 
Figure~\ref{ref:fit} shows the measured distributions of the observables used for 
the \dmt measurement overlaid with density estimates using \ttbar signal events with 
\dmt = \gevcc{-4} and \gevcc{0} and the full background model.  
The choice of \dmt = \gevcc{-4}~(solid line) gives better agreement with the data than 
that of \gevcc{0}~(dashed line). 

In conclusion, we examine the mass difference between $t$ and $\bar{t}$ quarks in 
the lepton+jets channel using data corresponding to an integrated luminosity of \invfb{5.6} from \ppbar 
collisions at $\sqrt{s}$~=~1.96~TeV. We measure the mass difference 
to be $\dmt = \mathrm{M}_{t} - \mathrm{M}_{\bar{t}} = \gevcc{\measStatSyst{-3.3}{1.4}{1.0}} = \gevcc{\measErr{-3.3}{1.7}}$. 
This result is consistent with CPT-symmetry expectation, $\dmt = \gevcc{0}$, with approximately 2$\sigma$ level deviations.  
%This result is approximately two standard deviations away from CPT-symmetry expectation, $\dmt = \gevcc{0}$, but 
%approximately two standard deviations away from the CPT hypothesis of zero mass difference. 
It is consistent with 
the recent result from the D0 Collaboration~\cite{d0result}, 
but is 2.2 times more precise. 
This is the most precise measurement of the mass difference between $t$ and $\bar{t}$ quarks.

\begin{acknowledgments}
We thank the Fermilab staff and the technical staffs of the participating institutions for their vital contributions. This work was supported by the U.S. Department of Energy and National Science Foundation; the Italian Istituto Nazionale di Fisica Nucleare; the Ministry of Education, Culture, Sports, Science and Technology of Japan; the Natural Sciences and Engineering Research Council of Canada; the National Science Council of the Republic of China; the Swiss National Science Foundation; the A.P. Sloan Foundation; the Bundesministerium f\"ur Bildung und Forschung, Germany; the Korean World Class University Program, the National Research Foundation of Korea; the Science and Technology Facilities Council and the Royal Society, UK; the Institut National de Physique Nucleaire et Physique des Particules/CNRS; the Russian Foundation for Basic Research; the Ministerio de Ciencia e Innovaci\'{o}n, and Programa Consolider-Ingenio 2010, Spain; the Slovak R\&D Agency; the Academy of Finland; and the Australian Research Council (ARC). 
\end{acknowledgments}


\begin{thebibliography}{99}
\bibitem{cdfco} We use a right-handed spherical coordinate system with the origin at the center of the detector. $\theta$ and $\phi$ are the polar and azimuthal angles, respectively.  The pseudo-rapidity is defined by $\eta = -\ln \tan(\theta/2)$.  The transverse momentum and energy are defined by $\pt = p \sin(\theta)$ and $\et = E \sin(\theta)$, respectively, where $p$ and $E$ are the momentum and energy of the particle.
\bibitem{met} The missing transverse energy, an imbalance of energy in the transverse plane of the detector, is defined by 
$\met = -\vert \sum_{towers}\et{\hat n_{T}} \vert$, where ${\hat n_{T}}$ is the unit vector normal to the beam and pointing to a given calorimeter tower and $\et$ is the transverse energy measured in that tower.
\bibitem{bbasym} We select a $b\bar{b}$ sample by requiring exactly two $b$-jets per event using a sample triggered on jet~($E_{T}>\gev{20}$). In addition, one $b$-jet is required to contain a soft muon from leptonic decays so that we can determine the charge of the $b$-quark associated with the jet. The energy scale of $b$ and $\bar{b}$ jets in the data is compared with di-jet MC events to estimate the uncertainty of $b$ and $\bar{b}$ jet energy scale independently. We perform PEs by varying the $b$ and $\bar{b}$ energy separately within their uncertainties.
\end{thebibliography}
\end{document}